\documentclass[12pt]{article} 
\usepackage{epsfig,amssymb,cite,multirow} 
\usepackage{amsmath}

\usepackage{amscd}
\usepackage{amsmath,graphicx}
\usepackage[matrix,arrow,curve]{xy}
\usepackage{graphicx}
\usepackage{verbatim}
\usepackage{epsf}
\usepackage{latexsym}
\usepackage{amsmath,amsfonts,amssymb,amsthm}
\usepackage{amsmath,amsthm}

\def\bseq{\begin{subequation}}  % = 1a 1b
\def\eseq{\end{subequation}}
\def\bsea{\begin{subeqnarray}}  % = 1.1a 1.1b
\def\esea{\end{subeqnarray}}
\newcommand{\bbox}{\lower.2ex\hbox{$\Box$}}

\newcommand{\beq}{\begin{equation}}
\newcommand{\eeq}{\end{equation}}
\newcommand{\bea}{\begin{eqnarray}}
\newcommand{\eea}{\end{eqnarray}}
\newcommand{\ena}{\end{eqnarray}}

\renewcommand{\)}{\right)}
\renewcommand{\[}{\left[}
\renewcommand{\]}{\right]}

\newcommand{\be}{\begin{equation}}
\newcommand{\ee}{\end{equation}}

\begin{document}
\setcounter{page}{0}
\begin{titlepage}
\titlepage
\begin{flushright}
%LPTENS-.........\\
UCSD-PTH-11-12\\
\end{flushright}
\begin{center}
\LARGE{\Huge Negative Refractive Index\\
in Hydrodynamical Systems}
\LARGE{\Huge  }
\end{center}
\vskip 1.5cm \centerline{{\bf Antonio Amariti$^{a}$\footnote{\tt amariti@physics.ucsd.edu},
Davide Forcella$^{b}$\footnote{\tt forcell@ulb.ac.be}, Alberto
Mariotti$^{c}$\footnote{\tt alberto.mariotti@vub.ac.be}
}}
\vskip 1cm
\footnotesize{

\begin{center}
$^a$Department of Physics, University of California\\
San Diego La Jolla, CA 92093-0354, USA
\\
\medskip
$^b$ Physique Th\'eorique et Math\'ematique and International Solvay Institutes\\
Universit\'e Libre de Bruxelles, C.P. 231, 1050Bruxelles, Belgium \\
\medskip
$^c$
Theoretische Natuurkunde, Vrije Universiteit Brussel \\
and
The International Solvay Institutes\\ 
Pleinlaan 2, B-1050 Brussels, Belgium\\
\end{center}}

\bigskip

\begin{abstract}
We discuss the presence of exotic electromagnetic phenomena in
systems with finite charge density which are described by hydrodynamics.
We show that such systems generically have negative refractive index
for low frequencies electromagnetic waves, 
i.e. the energy flux and the phase velocity of the wave propagate
in opposite directions.
We comment on possible phenomenological applications, focusing on the Quark Gluon Plasma.
\end{abstract}

\vfill
\begin{flushleft}
{\today}\\
\end{flushleft}
\end{titlepage}

\newpage

\tableofcontents

\section*{Introduction}
\addcontentsline{toc}{section}{Introduction}

Electromagnetic waves in continuous media have been a large field of investigation in the last years. 
Indeed light waves have an amazing behavior in a new class of artificial materials 
(metamaterials), created in laboratories about a decade ago \cite{Smith,Pendry}. 
One of the most attractive properties is the negative refractive index: namely the energy flux and the 
phase velocity of a wave packet propagate in opposite directions for a certain range of frequencies. 
This behavior led both to promising physical applications and to interesting theoretical developments.

In this paper we will discuss the existence of negative refraction in hydrodynamics. 
More precisely we show that \emph{negative refraction seems to be a generic phenomenon in homogeneous and 
isotropic systems that have a finite non zero charge density 
and that admit a description in term of hydrodynamical equations}, at least for certain range of their parameters.
This conclusion is valid both for normal and super conducting fluids, and for relativistic or non relativistic systems.  
The only necessary ingredients to reach this goal are linear hydrodynamics, classical
electrodynamics and linear response theory. 

Finally we speculate on the connection of our result with phenomenological and experimental applications.
Specifically  we concentrate on the possible presence of negative refraction in Quark Gluon Plasma, 
a strongly coupled plasma at finite charge density that should be created in high energy physics 
colliding experiments. 

Note that in \cite{Amariti:2010jw} it was already argued that  
negative refraction is an ubiquitous 
phenomenon in hydrodynamical charged systems, supporting the claim 
by the inspection of a strongly coupled theory where the transport coefficients
can be explicitly computed through the gauge/gravity correspondence.
 In \cite{Gao:2010ie} the same topic was investigated for 
an holographic superconductor in the probe limit, in \cite{Ge:2010yc} for 
a charged black hole in four dimensions, and in
 \cite{Bigazzi:2011it,Bigazzi:2011ut} for systems with D7 flavor branes.
In particular this last development seems promising for more
 realistic application to the Quark Gluon Plasma physics.
\\
\\
The structure of this paper is the following:
in Section (\ref{main}) we explain the physics of negative refraction by reviewing the
basic aspects of linear response theory and of electromagnetism in spatial dispersive media. In Section (\ref{sechydro}) 
we study the two points correlation function of the transverse current for a system described 
by hydrodynamical equations. In Section (\ref{secclaim}) we make our main claim: negative refraction is generic 
in charged liquids. In Section (\ref{secpheno}) we look for possible phenomenological examples and in particular we speculate 
about the QGP.

\section{Negative Refraction and Linear Response Theory}
\label{main}

The propagation of transverse polarized electromagnetic waves in a medium is usually 
described in terms of the refractive index $n$ defined as $n^2 = \epsilon \mu$,
where $\epsilon$ is the electric permittivity, while $\mu$ is the magnetic permeability. 
In real media they are both complex functions of the frequency $w$.
A medium is told to have negative refractive index \cite{negref,Agranovich} if for certain range of frequencies 
the phase velocity of a wave packet propagates in the opposite direction to the energy flux.

The phase velocity of a wave is defined as $\vec{v}_{ph}= 1/\hbox{Re}(n)\hat{k}$, where $\hat{k}$ is the unit vector in the direction of the wave vector $\vec{k}$,
while the direction of the energy flux, for not too dissipative and dispersive media,
is given by the Poynting vector, that for long wavelength is 
$\vec{S}=\hbox{Re}\(n/\mu\)\hat{k} |E_T|^2$ in the transverse channel, where $E_T$ is the 
component of the electric field transverse to $\vec{k}$.
In this setup negative refraction is equivalent to ask that the phase velocity and the Poynting vector are in opposite direction, namely $\hbox{Re}(n)<0$ 
and  $\hbox{Re}\(n/\mu\)>0$.
It is possible to show \cite{Depine} that in passive media
 the energy flow and the phase velocity are
opposite if and only if 
\begin{equation}
\label{delfino}
n_{DL}(w) = |\epsilon(w)| \hbox{Re}(\mu(w)) +|\mu(w)|
\hbox{Re}(\epsilon(w)) 
\end{equation}
is negative. 
Hence negative refraction is equivalent to $n_{DL}(w)<0$ for certain range of frequencies $w$.
Our task is to find a general class of media for which this condition is satisfied.
This aim can be reached only after the computation of the response functions 
$\epsilon$ and $\mu$.

These two quantities can be obtained by studying the retarded two point
correlation function of the 
electromagnetic current in the medium. 
Here we concentrate on the propagation of transverse waves in
homogeneous and isotropic media that admit an hydrodynamical description 
for a certain range of the physical parameters. 
Hydrodynamical media are able to transmit external perturbation quite easily, implying that their response functions 
generically depends from the wave vector $k$ and not only from the frequency $w$. 
This property is called spatial dispersion.

The macroscopic electromagnetism in media with spatial dispersion is defined in term
of the three fields 
$D$, $E$, $B$, and the relation $D_i= \epsilon_{ij}(w,k) E_j$. These fields
satisfy the macroscopic 
Maxwell equations and the tensor function $\epsilon_{ij}(w,k)$ describes the
linear response of the 
medium \cite{Landau}. For an isotropic medium: 
$\epsilon_{ij}(w,k)= \epsilon_T (w,k) P^T_{ij} + \epsilon_L(w,k)P^L_{ij}$ where $P^T_{ij}$ ans $P^L_{ij}$ are the transverse 
and longitudinal projectors\footnote{$P^T_{ij}= \delta_{ij} - \frac{k_i k_j}{|k|^2}$ and $P^L_{ij}=\frac{k_i k_j}{|k|^2}$.} with respect to the wave vector $k$, while $\epsilon_T$ and $\epsilon_L$ are two scalar functions of $k$ and $w$.
 
The propagation of the transverse part of the electromagnetic field requires that the dispersion relation $\epsilon_T(w,k)= k^2 / w^2$ is satisfied. 
If we probe the medium with an electromagnetic field with large enough wave-length we can expand $\epsilon_T$ at
the second order in $k$: $\epsilon_T(w,k) = \epsilon(w) + \frac{k^2}{w^2}
\left( 1 - \frac{1}{\mu(w)} \right) \,$, where $\epsilon(w)$ the electric permettivity, and $\mu(w)$ is the effective magnetic
permeability\footnote{It is important to observe that the effective magnetic permeability 
we use here is different from the usual magnetic permeability \cite{Landau}, but it is the right quantity to describe the propagation of transverse waves in spatially dispersive media 
\cite{Agranovich}.}.

The electric permittivity and the magnetic permeability can be obtained from linear
response theory \cite{Dressel}. 
In the linear response theory the electromagnetic current $J_i$ is proportional to
the vector potential $A_j$, by $J_i= \hbox{ }G_{ij} A_j$, where $G_{ij}$ is the retarded correlator of the currents in the medium.

We can decompose $G_{ij}$ in its transverse and longitudinal part, using the projectors $P^T_{ij}$ and $P^L_{ij}$, and obtain the relation
 \footnote{Here we follow the opposite convention of \cite{Amariti:2010jw} for the sign of the Green function 
 to be consistent with the conventions common in the literature (es. \cite{Foster}).
This changes the sign of the $q^2$ contribution in $\epsilon_T$.}: 
$\epsilon_T(w,k)= 1+ \frac{4 \pi}{w^2}\hbox{  } \hbox{ }G_T(w,k) \,$.
Expanding $G_T$ to second order in $k$: $G_T(w,k)= G_T^{(0)}(w)+ k^2 
G_T^{(2)}(w)$,
we find the electric permittivity and effective magnetic
permeability
\begin{eqnarray}
\label{fighii}
&&\epsilon(w)= 1+ \frac{4 \pi}{w^2} \hbox{  } G_T^{(0)}(w)
\nonumber \\
&&\mu(w)=\frac{1}{1- 4 \pi  \hbox{  } G_T^{(2)}(w)}
\end{eqnarray}

In summary, for long enough wavelength, the propagation of the transverse part of the electric field
in the medium can be described in terms of the first orders expansion in $k$ of the retarded 
correlator of transverse currents.

\section{Considerations from Hydrodynamics}
\label{sechydro}

In this section we want to discuss some generic properties of the propagation of electromagnetic waves in a charged hydrodynamical system.
By hydrodynamical system we mean every system that, for a certain range of its parameters, admits a description in term of hydrodynamical equations for large enough distance and long enough time.
Hydrodynamics is indeed the effective theory describing the real-time dynamics of a system at scales larger than the mean free path and time scales larger than the mean free time 
of the microscopic elements composing the system.
At these scales only a finite number of degrees of freedom survives: densities of conserved charges and phases of order parameters. 

Our main interest is to determine the generic form of the retarded two points correlation function of the transverse part of the electromagnetic current.
This result will be obtained by combining hydrodynamical equations and linear dynamics response theory.

For simplicity we focus on the case of a non-relativistic fluids in the normal phase, but, as we will comment at the end of the section, the results can be generalized to other interesting cases.
A generic non-relativistic fluid can be described by the Navier-Stokes equations. We just consider the linearized version of the Navier-Stokes equations because we are interested in linear response .
They essentially come from the conservation of the particle density, energy density and the momentum density in addiction to
 some constitutive relations and thermodynamic relations.
The conservation for the density $n$ takes the form: $\partial_t n + \nabla_i p_i/m=0$, where $p_i$ is the momentum density and $m$ the characteristic mass of the particles. 
If the system has another conserved charge associated to some U(1) symmetry, that we will reinterpret later as the electromagnetic charge, and if every particle in the system has charge $e$ under this symmetry, than the conservation equation for the charge density of this symmetry $\rho= e n$ is just  proportional to the particle density conservation:
$\partial_t \rho + \nabla_i j_i=0$, with $j_i$ the current associated to the conserved charge density. 
This observation suggests that the properties of the transverse part of the current density $j_i$ are related to the 
dynamics of the transverse part of the momentum density $p_i$. 
 
 In the Navier-Stokes system also the momentum density is conserved: $\partial_t p_i + \nabla_i \tau_{il}=0$, where $\tau_{il}$ is the stress tensor. The momentum density can be divided into its longitudinal $\nabla \wedge p^L=0$  and transverse part $\nabla \cdotp p^T=0$.
 We are interested to the transverse part $p^T$ only.  
 The equation describing the dynamics of the transverse momentum decouples from all the other equations \cite{KM,Foster}:
 \begin{equation}\label{difeqpT}
 (\partial_t - \frac{\eta}{nm} \nabla^2 ) p^T = 0
 \end{equation}
where $\eta$ is the shear viscosity of the system.
The decoupling of the transverse momentum largely simplifies our analysis and it is valid also for relativistic, non-relativistic, superconducting or normal fluids. Indeed temperature and chemical potential inhomogeneities, and the derivative of the phase of the order parameter for the superfluid phase, only couple to the longitudinal part of the momentum density\footnote{For a charged fluid the electromagnetic field couples to 
the particle system and, at the linear level, for the case we are interested in, one has to 
solve the coupled equations:$(\partial_t - \frac{\eta}{nm} \nabla^2 ) p^T = e n \partial_t A^T$, 
$(\partial_t^2-\nabla^2)A^T= J^T$, with the constraint $J^T= (ep^T - e^2n A^T)/m$. However, 
in the limit of small coupling $e$, it is possible to show that this effect gives a subleading contribution
to the retarded correlator of the transverse currents.
 It is important to observe that the order of limits matters; in all the paper we
 consider small $e$ before then small $k$ and then for last small $w$.}.
Equation (\ref{difeqpT}) is solved by Fourier transforming  in the space variable $x$ 
and by Laplace transforming in the time variable $t$. After we
project on the real frequencies we have
 \begin{equation}\label{Hydropi}
p^T(w,k)=\frac{p^T(t=0,k)}{ -i w + \frac{\eta}{nm} k^2}
 \end{equation}
where $p^T(t=0,k)$ is the Fourier transform of the momentum density at time $t=0$.
This solution describes the evolution of the transverse part of the momentum density: the system is displaced from its equilibrium configuration by an external force and then it is left to evolve freely in time. The same information is contained in the long wave, low frequencies behavior of the two point retarded correlation function of the transverse momentum $G_{p^T}(w,k)$. The correlation functions can be typically computed using the linear response theory. Linear hydrodynamics and linear response theory at long time and large scale are two different descriptions for the same physics, and they must coincide \cite{KM,Foster}.  

Indeed the evolution of a degree of freedom $\mathcal{O}(t,x)$ of the system due to the perturbation of the equilibrium state with an external field $\mathcal{\phi}(t,x)$, can be described solving the initial value problem associated to the hydrodynamical equation of the system: $\mathcal{O}_H(w,k)$, or computing the two points retarded correlation function $G_{\mathcal{O}}(w,k)$ of the observable $\mathcal{O}$ according to: $\mathcal{O}_{LR}(w,k)=G_{\mathcal{O}}(w,k) \phi(w,k)$.
A very general result from linear response theory says that for adiabatically applied perturbation we have:
\begin{equation}\label{fundpi}
\mathcal{O}_{LR}(w,k)=\frac{G_{\mathcal{O}}(w,k) - G_{\mathcal{O}}(k)}{i w ~ G_{\mathcal{O}}(k)} \mathcal{O}(t=0,k)
\end{equation}
where $G_{\mathcal{O}}(k)$ is the $w\rightarrow 0$ limit of $G_{\mathcal{O}}(w,k)$.
This formula eliminates the explicit dependence on the external field and it describes the relaxation process as an initial value problem.
For small enough frequencies and wave vectors: $\mathcal{O}_{LR}(w,k)=\mathcal{O}_{H}(w,k)$ or in other words we can use hydrodynamics to infer the long time and large distance behavior of the retarded correlator, in the regime in which it is usually difficult to compute it directly, due to the collective interactions of the system.
 
From this fundamental result we can now study the case of the transverse momentum.
Equating the expression (\ref{fundpi}) for $\mathcal{O}=p^T$ with the hydrodynamical result (\ref{Hydropi}), we obtain: 
\begin{equation}\label{corrpiT}
G_{p^T}(w,k) = \frac{ i w m n }{ - i w + \frac{\eta}{m n } k^2 } + mn 
\end{equation}
where we took the $k\rightarrow 0$ limit and we used equilibrium statistical mechanics to show that $G_{p^T}(0,0)= mn$ \cite{Foster}.

Starting from (\ref{corrpiT}), we  compute the retarded correlator of the transverse current: $G_{J^T}(w,k)$. The induced transverse current $J^T$ is the response of the system to an external electromagnetic field $A_{ext}^{\mu}$. The standard interacting hamiltonian density is: $\delta \mathcal{H}= -J_{\mu} A_{ext}^{\mu}$.

It is important to stress that in presence of an electromagnetic field the transverse velocity density field of the system is no more $p^T/m$ but $(p^T -  e n A^T)/m $, due to the minimal coupling of the system to the vector potential.
The transverse current is proportional to the transverse field velocity $v^T$ of the system \cite{Foster,KM,Herzog,Hartnoll:2007ih} and in particular:
\begin{equation}\label{emcur}
J^T= e n v^T= \frac{e}{m}p^T -  \frac{e^2 n}{m}A^T
\end{equation}
The Green function $G_{J^T}(w,k)$ is obtained by assuming that all the electromagnetic fields in the system are external;
explicitly we put $A=A_{ext}$ everywhere \cite{Foster}. This hypothesis may appear too strong and it amounts to treat 
even the internally generated electromagnetic field as an external perturbation; in this case the equation (\ref{difeqpT}) is exact. We believe that this assumption can be justified in various systems and regimes. We take it as a working hypothesis, and we will justify it in the following, when we will discuss possible phenomenological applications.
With this approximation the problem simplifies and the interacting hamiltonian density becomes $\delta \mathcal{H}= -j^T A_{ext}^T$, where $j^T=p^T e/m$ is the translational current density, and: 
\begin{equation}
j^T(w,k)=G_{j^T}(w,k)A^T_{ext}=\frac{e^2}{m^2} G_{p^T}(w,k)A^T_{ext}
\end{equation}
Using equations (\ref{emcur}), (\ref{corrpiT}) and the fact that $J^T=G_{J^T} A^T_{ext}$ we obtain: 
\begin{equation}\label{retJTsf}
G_{J^T}(w,k) = \frac{ i w \frac{e^2 n}{m}}{ - i w + \frac{\eta}{nm} k^2 } 
\end{equation}
This result is valid for a normal, non-relativistic fluid. 
Analogously a similar result also for relativistic and/or superfluid systems can be deduced.
Indeed from linearized hydrodynamics and linear response theory the generic form of the retarded correlator for the transverse part of the electromagnetic current is:
\begin{equation}\label{genG}
G_{J^T}(w,k) = \frac{ i w \mathcal{B}}{ - i w + \mathcal{D} k^2 } - \mathcal{C}
\end{equation}
with $\mathcal{B}$, $\mathcal{C}$ and $\mathcal{D}$ three real positive constants depending on the particular system. The $\mathcal{C}$ constant is a contribution coming from the superfluid phase and it is zero in the normal phase. An important example that we analyze in the following section is a relativistic charged fluid for which $\mathcal{B}=\rho^2/(\epsilon + P ) $ and $\mathcal{D}= \eta/(\epsilon + P)$, where $\rho$ is the charge density, $\epsilon$ the energy density and $P$ the pressure. For completeness we give also the explicit expressions for the superconducting phase: in the non-relativistic case $\mathcal{B}=e^2 n_n/m $, $\mathcal{D}= \eta/m n_n$ and $\mathcal{C}= e^2 n_s/m$, while in the relativistic case $\mathcal{B}=\rho^2_n/( \epsilon + P -\rho_s \mu) $, $\mathcal{D}= \eta/( \epsilon + P -\rho_s \mu)$ and $\mathcal{C}= \rho_s \rho_n/( \epsilon + P -\rho_s \mu)$, where $n_n$ is the particle density of the normal component of the fluid, while $n_s$ is the particle density of the superfluid component, $\rho_n$, $\rho_s$ is the normal and superconducting charge density and $\mu$ is the electromagnetic chemical potential.
 
\section{Hydrodynamics and Negative Refraction}
\label{secclaim}

From the generic form of (\ref{genG}) we can obtain $G^{(0)}(w)$ and $G^{(2)}(w)$ and hence the generic form of $\epsilon(w)$ and $\mu(w)$:
\begin{equation}\label{epsmu}
\epsilon(w)= 1 - \frac{4\pi}{w^2}\left( \mathcal{B} + \mathcal{C} \right), \qquad \mu(w)= \frac{1}{1- 4 \pi \frac{ i \mathcal{B} \mathcal{D} }{ w}} 
\end{equation}
From the equations (\ref{epsmu}) we observe that for small enough frequencies $n_{DL}(w)$ is negative. 
In particular $n_{DL}(w)<0$ for $w< 2 \sqrt{\pi (\mathcal{B}+ \mathcal{C})}$.
Hence we can conclude that \emph{an isotropic and homogeneous system, in local thermal and mechanical equilibrium, invariant under relativistic or non-relativistic space time transformations, at finite charge density and in the normal or superconducting phase, described for long enough time and large enough distance by 
hydrodynamical equations, has negative refractive index for small enough frequencies}.

As we explained we computed the response functions
after expanding $G_T({w,k}) $ for small $k$. This expansion is valid if
$|k^2| \ll  w/D$, otherwise the effects of spatial dispersion are very strong and 
the whole $k$ dependence has to be taken into account.  This gives origin to multiple solutions for 
the dispersion relation $n^2(w) = \epsilon(w) \mu(w)$,
the so-called additional light waves (ALWs) \cite{Pekar,ALW}. 
In the rest of this note we work in the regime $|k^2| \ll w/D$ neglecting the existence of ALWs.
In this regime the large wavelength constraint becomes $|n^2(w)| \ll \frac{1}{w D}$, and it imposes a constraint on the frequency range that we can explore. Moreover, because we are always working in the hydrodynamical limit $w$ and $k$ must be small compared to the temperature and the chemical potential of the system.  This requirement introduce an additional constraint on the allowed frequency range. 

Another interesting quantity to compute is the ratio $\Big|\frac{Re\[n(w)\]}{Im\[ n(w) \]} \Big|$.  If it is larger than one means that the propagation of the electromagnetic waves dominates over the dissipation. In our example unfortunately this ratio is smaller than one in the frequency range in which we have negative refraction, meaning that the dissipation is an important phenomenon. However this seems to be the normal behavior for isotropic metamaterials that experience negative refraction. In the actual experiments and simulations this problem is typically solved by introducing anisotropies or by
considering superconducting materials. We 
discuss some of these developments in \cite{siani},
 while here we concentrate on the easiest example of a non superconducting and isotropic hydrodynamical system.
It is important to observe that, even if the final result shows relevant dissipation effects, 
the procedure that we have just described  is reliable in the regime of small spatial dispersion. 

To conclude it is important to observe that  in the hydrodynamical derivation  
we assumed that the external field $A_{ext}$ is the dominating one.  
This analysis ignored the contribution of the  induced  internal field. 
By taking into account its contribution some constraints are imposed on the validity of the 
$q^2$ expansion.  This requirement enforces another lower bound on the frequency.
We checked that there are regimes of parameters in which this bound doesn't 
affects the whole region of negative refraction. Anyway we leave deeper analysis for further studies.

\section{Phenomenology}
\label{secpheno}

The arguments we gave in the previous sections bring us to the conclusion that negative refraction is a quite common phenomenon in charged hydrodynamical systems. 
It becomes important to understand the range of applicability of our conclusion to physically observable systems.

The main properties of the system we are describing is its finite charged density and the fact that it is in local thermo-mechanical equilibrium. 
These two facts together lead to important constraints. 
Indeed the presence of non zero charge density generates forces\footnote{It is however important to observe that 
only the longitudinal part of the electric field $E^L$ is influenced by the presence of a finite charge density $\rho$.} 
that will eventually repel its constituents,  showing that the system is actually out of equilibrium.   
This conclusion is false for infinite or three dimensional periodic systems, that are actually what our formalism is describing. 
On the contrary side, infinite or three dimensional periodic systems do not exist.
For this reason we need to admit that our model describe only the bulk properties of a finite sample, of which we disregard the boundary effects. 
However the existence of finite systems at finite charge density and in local equilibrium is not obvious.

A possible way out is to consider the systems we are describing as part of a globally neutral system. 
An intriguing example is to consider the electrons of a metal as a charged fluid. In the approximation in which the lattice ions in the metal 
provide a static homogeneous charge distribution, the entire system is globally neutral and in mechanical equilibrium, and the system should have 
negative refraction for certain range of parameters
\footnote{We thanks Jan Zaanen for discussions related to this point.}.

In this paper we will not pursue this intriguing line of research related to solid state physics, but we will instead 
focus on another interesting example, related to high energy physics.
Indeed another way out is to think that we are describing the bulk dynamics of 
the whole finite system, in which there exist other forces that compensate for the 
electromagnetic repulsion.
A natural candidate for this line of thinking is the QGP (Quark Gluon Plasma): 
a strongly coupled plasma that is believed to form in the high energy collisions of 
ions inside the particle accelerators.  
This plasma is globally charged due to charge conservation.
Moreover it is believed to be in local thermo-mechanical equilibrium, at least for a small fraction of time,  
and to be well approximated by hydrodynamics.
Finally the 
QGP dynamics is dominated by the strong interaction among its constituents,
so it seems justified to neglect the internal electromagnetic field.

These two examples are clearly in contrast with two common lores:  there are no natural materials with negative refraction, 
the QGP is insensitive to the electromagnetic interaction; 
however the simplicity and generality of the effects we described in the previous sections suggest 
that it is maybe worthwhile to investigate more along this direction.

\subsection{A calculable example}
In this section we perform
a more quantitative analysis of negative refraction in 
hydrodynamical systems and we speculate about its application to the
QGP.
In order to make a concrete prediction, we need to specify
the hydrodynamical coefficients entering the green function 
and then the electromagnetic parameters (\ref{fighii}).

Recent advances in high energy physics offer explicit models that 
belong to the universality class 
we have just discussed. In these physical systems, exact values of the
thermodynamical parameters and transport 
coefficients are already at our disposal.
These models are believed to describe some aspects of the strongly 
coupled plasma (quark gluon plasma) that is supposed to form in 
particle colliders
and is outside the regime of perturbative QCD\footnote{
Despite some of their successes 
it is important to stress that 
these models describe a different theory than QCD.
For this reason we can just use them to provide an
approximate estimate of physical quantities. }.
%like RHIC or LHC. 
They are based on the gauge-gravity correspondence (AdS/CFT),
i.e. a relation between a strongly coupled $d$ dimensional field theory and 
a gravity theory in $d+1$ dimensions.
Here we skip the details of the correspondence, referring the interested reader 
to the huge literature \cite{Aharony}, 
and we straightforwardly move to the 
results\footnote{Note that in the context of the AdS/CFT correspondence, 
the electromagnetic field is by construction an external perturbation.}.

%Indeed 
In \cite{Amariti:2010jw},
using the results of \cite{korea},
we 
computed the electromagnetic response function
of a four dimensional system dual
to an $AdS_5$ charged black hole,
and 
we found negative refraction at low frequencies.
The green function for that setup is
%e' meno la green function che abbiamo sempre usato noi
\be
\label{greenads}
G_{J^T}(w,k)= \frac{i w \mathcal{B}}{-i w + \mathcal{D} k^2} + i w \mathcal{R} +\mathcal{N} (w^2-k^2)
\ee
where the first term is the leading one we derived with pure hydrodynamics (\ref{genG}),
whereas the extra two terms are specific of the AdS/CFT computation. They are subleading with respect to the hydrodynamical 
ones, but they give $O(1/w)$ and constant contribution
to the electromagnetic response function $\epsilon_T$, changing in this way the quantitative results.
The permittivity and the permeability then read
\be
\label{epsmuads}
\epsilon(w)=1+\frac{4 \pi}{w^2} \left(-\mathcal{B} + i w \mathcal{R} + w^2 \mathcal{N}  \right)
\qquad
\mu(w)=\left(1-4 \pi (\frac{i \mathcal{B} \mathcal{D} }{w} - \mathcal{N})
\right)^{-1}
\ee
The various coefficients are
\be
\label{coeffads}
\mathcal{B}= \frac{\rho^2}{\epsilon+P} \qquad \mathcal{D}=\frac{\eta}{\epsilon+P}
\qquad \mathcal{R}=
8 \pi^2 T^2 
\frac{\eta}{(\epsilon+P)^2}
\left(
\frac{\rho}{\mu}
\right)^2
\qquad \mathcal{N}= \text{constant} 
\ee
The charge density, energy density and pressure, and shear viscosity are 
determined to be
\bea
&&
\rho= e
\frac{\sqrt{3} N_c^2 \left(\pi  T+\frac{\sqrt{3 \pi ^2 T^2+ \mu ^2}}{\sqrt{3}}\right)^3 \sqrt{ \mu ^2+2 \pi  T \left(3 \pi  T-\sqrt{3} \sqrt{3 \pi ^2 T^2+ \mu ^2}\right)}}{32 \pi ^2  \mu }
\nonumber \\
&&
\epsilon + P=
\frac{N_c^2 \left(\pi  T+\frac{\sqrt{3 \pi ^2 T^2+ \mu ^2}}{\sqrt{3}}\right)^4 \left(3  \mu ^2+4 \pi T \left(3 \pi T-\sqrt{3} \sqrt{3 \pi ^2 T^2+ \mu ^2}\right)\right)}{32 \pi ^2  \mu ^2}
\label{termoads}
\\
 &&
  \eta=\frac{N_c^2 \left(\pi  T+\frac{\sqrt{3 \pi ^2 T^2+ \mu ^2}}{\sqrt{3}}\right)^3}{64 \pi ^2} \nonumber
\eea
Note that the only free parameters in the permittivity and in the 
permeability are the temperature $T$, the electromagnetic chemical
potential $\mu$ and the dimensionless parameters $N_c$. 
The electric charge $e$
is fixed to its classical value in natural units $e=0.3028$.
The parameter $\mathcal{N}$ comes from the UV frequencies through 
the renormalization procedure, and it is fixed by
requiring that the permittivity approaches unity\footnote{Namely for very high frequencies the electromagnetic waves is supposed to probe the vacuum.} for $w \to \infty$.
The parameter $N_c$
is the rank of the $SU(N_c)$ gauge group of the underlying
strongly coupled four dimensional field theory. 
The expressions (\ref{epsmuads})
are valid in the hydrodynamical regime, where both the frequency
and the wave vector are small compared to the other scales of the theory, namely the
temperature and/or the chemical potential.

We can proceed by plotting some numerical result for the critical
frequency below which we predict negative refraction.
Since we are interested in quark gluon plasma physics, we can 
conventionally set $N_c=3$ as the
number of colors of QCD, from now on.
In Figure 1 (blue line) we plot the ratio of the critical frequency $w_c$ 
over the temperature
as a function of the ratio between the chemical potential and
the temperature. We see that the region of negative
refraction grows with $\mu/T$.
However our analysis does not cover the entire region 
of frequencies below $w_c$ due to the constraints
explained in the previous section. The
allowed range is restricted to $w_0/T<w/T<w_c/T$.
In the Figure 1 we plot also the minimal allowed frequency (red line) $w_0/T$.
The region where our approximations are valid and which exhibits
negative refraction is the one in between the two lines.

\begin{figure}
\begin{center}
\includegraphics[width=10cm]{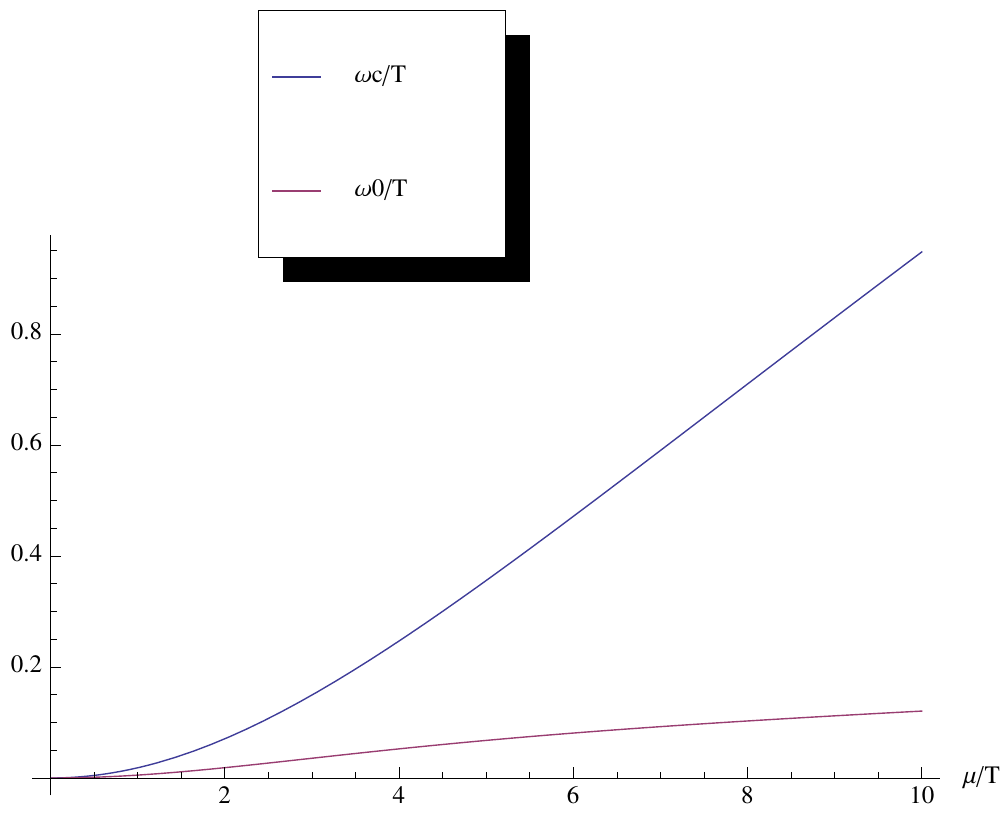}
\caption{Critical frequency $w_c/T$ and the minimal allowed frequency $w_0/T$ as a function of the chemical potential $\mu/T$.}
\label{frequ}
\end{center}
\end{figure}

Our results are supposed to describe some of the electromagnetic 
bulk properties of the system. 
However, for this to be reliable,
the characteristic wavelength of negative refraction must be
smaller than the typical length scale 
of the experiment.

For this reason 
we have to analyze the wave length
$\lambda(w)=\frac{2 \pi}{\text{Re}[n(w)] \, w}$.
Note that the real value of the refractive index
Re$[n(w)]=$Re$[\sqrt{\epsilon \mu}]$ 
vanishes at $w_c$ and increases for smaller frequencies \cite{Amariti:2010jw}.
Thus the wavelength $\lambda(w)$ diverges at $w_c$ and 
monotonically decreases for smaller frequencies.
The minimal length that can be probed by the negative
refraction is then set by $\lambda(w_0)\equiv\lambda_0$.
In Figure 2 we plot $\lambda_0 T$ as a function of $\mu/T$.
It is negative to stress that there is negative refraction.

\begin{figure}
\begin{center}
\includegraphics[width=10cm]{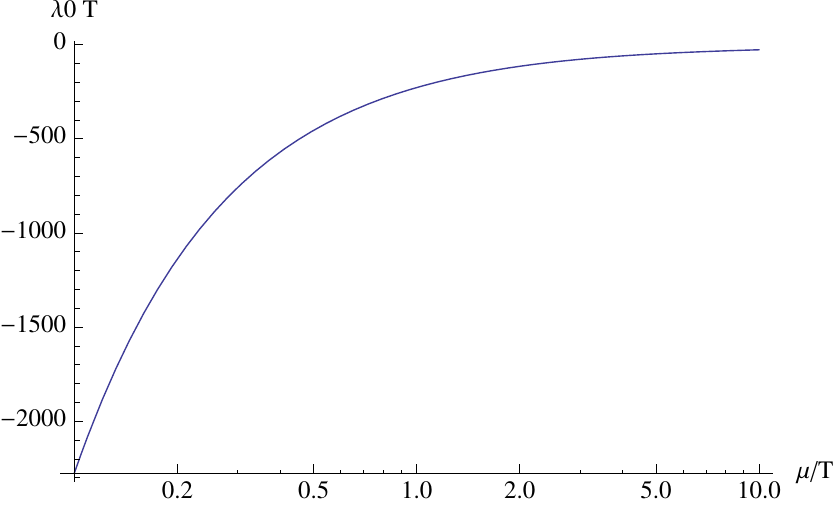}
\caption{Minimal wavelength $\lambda_0/T$ with negative refraction as a function of the chemical potential over the temperature.}
\label{lamb}
\end{center}
\end{figure}

For a generic system, 
given the temperature and the electrical chemical
potential, one can extract from the 
Figures 1 and 2 the frequency interval and the smaller wavelength 
where we expect negative refraction.

Observe that even if the analysis has been made 
in the hydrodynamical approximation, the actual physical values
of the frequency
can be large, if the temperature and/or the chemical
potential are large. In the following,
for instance, we find $w_c \sim 10^{21}$ Hertz
for the QGP.

\subsection{QGP and Negative Refraction}

Here we give some numerical results about negative refraction
in the QGP.
In the description of the physics of ion collisions (like nuclei of $Au$ or $Pb$),
the independent parameters are the temperature and the baryonic
chemical potential $\mu_B$. 
The phase diagram of QCD as a function of these variables 
is not completely explored. 
The common lore is that QCD is a plasma for
temperature around  
$T_c \simeq 150 - 170 MeV$ and energy density 
$\epsilon_c \simeq O(GeV/fm^3)$.
Useful references for the QGP physics and the experiments are
\cite{Andronic,Redlich,Mateos}.

We are interested in identifying the typical 
frequencies and wavelength where
we expect negative refraction in the QGP.
For this purpose we need the values
of $T$ and of the electromagnetic chemical potential $\mu$. 
We consider the simplifying assumption that
the sample of colliding ions has the same number
of protons and neutrons, i.e. we neglect the total isospin charge.
In this approximation the electromagnetic chemical potential 
results 1/2 the baryonic
one.

The estimated temperature and baryonic chemical potential
 vary among the different experiments (e.g. AGS, SPS, RHIC)
and depending on the data that are taken into account.
For simplicity we can consider three benchmark points 
for the values of $T$ and $\mu_B$,
which correspond approximately to the values at the freeze 
out\footnote{The freeze out curve in the $T,\mu_B$ plane 
is determined
by the fact that the inelastic collisions cease,
and in some cases it corresponds to
the critical line of the QCD phase transition between hadrons and the QGP.}
for the three experiments AGS, SPS and RHIC.
Modifying the ratio and/or the critical temperature of few percents
 does not change significantly our conclusions.

Using (\ref{termoads}),(\ref{coeffads}) and (\ref{epsmuads}) we compute the thermodynamical
quantities for the three benchmark points, 
and the frequency range and the minimal
wavelength where we expect negative refraction:

\begin{center}
\begin{tabular}{|c|c|c|c|}
\hline
& $\sim$ AGS & $\sim$ SPS & $\sim$ RHIC \\
\hline
$T$ & 125 MeV & 148 MeV &  177 MeV \\
\hline
$\mu=\frac{\mu_B}{2}$ & 270 MeV & 200 MeV &  14.5 MeV \\
\hline
$\rho/e$ &  $0.67 /fm^3$ & $0.66 /fm^3$ & $0.066 /fm^3$ \\
\hline
$\epsilon$ & $1.32 GeV /fm^3$ & $2.29 GeV/ fm^3$  & $ 4.28 GeV/fm^3$ \\
\hline
$[ w_0, w_{c} ]$ & $ [4.02, 15.5] \times 10^{21} Hz$ & $ [2.07, 7.46]  \times 10^{21} Hz$  & $
[9.79, 33.6] \times 10^{18} Hz$ \\
\hline
$\lambda_0$ & 168 fm
& 225 fm & 3100 fm\\
\hline
\end{tabular}
\end{center}
Unfortunately the typical wavelength is larger than the typical dimension
of the sample of quark gluon plasma, which is of the order of few fm,
so negative refraction
can probably not be tested in the actual collider experiments.
The probed wavelength becomes smaller with increasing $\mu/T$, but
in order to reach the fm, the chemical potential should be very large,
$\mu/T \sim 100$, so probably in region beyond the plasma phase of QCD.

In this section we didn't want to give precise numerical results,
but just point out that high energy physics can offer examples of 
finite charged systems where electromagnetic 
exotic phenomena could appear.
It would be appealing to 
use models which are better approximation of QCD
to give more quantitative predictions on the electromagnetic
properties of QGP.

\section{Conclusions}

In this paper we have shown that
systems at finite charge density that can be described by hydrodynamics
have negative refractive index for low frequencies electromagnetic waves.
We supported this claim computing the general properties 
of the  linear response of the system to an external electromagnetic field.
We hope that our discussion could stimulate interest on this faschinating 
topic both at theoretical and phenomenological level.
Indeed the negative refraction in the low
frequency regime is generic in both relativistic and non relativistic systems, both in a normal phase or in a superfluid one.
We restrict our phenomenological 
interest to the QGP but in principle the same observation 
can be valid in many different cases.

There are many directions to pursue the research on this subject.

At theoretical level it would be nice to generalize this effect to more
realistic setups,
relaxing some of the constraints,
to better understand the robustness of our result.
Indeed, while the requirement of a finite charge density 
seems intrinsic in our claim, 
the other hypothesis of homogeneity and isotropy may, in principle, be relaxed. 
Moreover, introducing boundaries in our analysis 
would also be very interesting for concrete
applications.

At the level of AdS/CFT it would be very nice to use better approximations of
the QGP to see if this electromagnetic effect could have relevant high energy  
physics phenomenology.
Moreover some metamaterial, the so called stereometamaterials \cite{Stereo}
seems to be intrinsically strongly coupled and studying the non hydrodynamical
modes in AdS/CFT could give new results related to the propagation 
of electromagnetic waves in strongly coupled plasmas.

Finally, at the phenomenological level, it would be very nice to understand under which 
approximations some common solid states systems can be described as charged
fluid and show in this way negative refraction.
Conducting electrons in metal
could reveal nice surprises.

\section*{Acknowledgments}
It is a great pleasure to acknowledge    D.~Carati, J.~Casalderrey-Solana,
A.~Cotrone, C.~Destri, J.~ Evslin, F.~Ferrari, 
V. Ginis,  S.~Hartnoll, G.~Kozyreff,  D.~Mateos, 
D. Musso, J.~ B.~ Pendry,
M.~Tlidi,  I. Veretennicoff and 
J.~Zaanen for nice discussions 
and useful comments on the draft. 
\\
A.A. is supported by UCSD grant DOE-FG03-97ER40546; the work
of D.F. is partially supported by IISN - Belgium (convention 4.4514.08), by the Belgian
Federal Science Policy Office through the Interuniversity Attraction Pole P6/11 and by
the ÒCommunaut«e Francaise de BelgiqueÓ through the ARC program;
A. M. is a Postdoctoral researcher of FWO-Vlaanderen. A. M. is also supported in part
by the Belgian Federal Science Policy Office through the Interuniversity Attraction Pole
IAP VI/11 and by FWO-Vlaanderen through project G.0114.10N.
\\
D.F. would like to thank for the kind hospitality the organizers of VI Avogadro
Meeting on Strings, Supergravity and Gauge theories, in GGI, Firenze, the
Theoretical Physics Group in 'Ecole Polytechnique, Paris, the Department of Particle
Physics and 
Instituto Gallego de Fisica de Altas Energ'as (IGFAE) de la Facultad de Fisica de la 
Universidad de Santiago de Compostela, the Department "Galileo Galilei" of the
Padova University, the Theoretical Physics Group in the University of Milano-Bicocca
and the Department of Theoretical Physics in Turin University, where part of the
work has been done.


\begin{thebibliography}{99}

\bibitem{Smith}
D. R. Smith, W. J. Padilla, J. Willie, D. C. Vier, S. C. Nemat-Nasser. and S. Schultz,
``Composite Medium with Simultaneously Negative Permeability and Permittivity,"
 Physical Review Letters, vol. 84 no. 18 (2000), pp. 4184 - 4187 
 
 \bibitem{Pendry}
J. B. Pendry
``Negative Refraction Makes a Perfect Lens,"
Phys. Rev. Lett. 85, 3966�3969 (2000) 

\bibitem{Amariti:2010jw}
  A.~Amariti, D.~Forcella, A.~Mariotti, G.~Policastro,
  ``Holographic Optics and Negative Refractive Index,''
  [arXiv:1006.5714 [hep-th]].




  %\cite{Gao:2010ie}
\bibitem{Gao:2010ie}
  X.~Gao, H.~Zhang,
  ``Refractive index in holographic superconductors,''
  JHEP {\bf 1008 } (2010)  075.
  [arXiv:1008.0720 [hep-th]].

%\cite{Ge:2010yc}
\bibitem{Ge:2010yc}
  X.~-H.~Ge, K.~Jo, S.~-J.~Sin,
  ``Hydrodynamics of RN AdS$_4$ black hole and Holographic Optics,''
  JHEP {\bf 1103 } (2011)  104.
  [arXiv:1012.2515 [hep-th]].

%\cite{Bigazzi:2011it}
\bibitem{Bigazzi:2011it}
  F.~Bigazzi, A.~L.~Cotrone, J.~Mas, D.~Mayerson, J.~Tarrio,
  ``D3-D7 Quark-Gluon Plasmas at Finite Baryon Density,''
  JHEP {\bf 1104 } (2011)  060.
  [arXiv:1101.3560 [hep-th]].


%\cite{Bigazzi:2011ut}
\bibitem{Bigazzi:2011ut}
  F.~Bigazzi, A.~L.~Cotrone, D.~Mayerson, A.~Paredes, J.~Tarrio,
  ``Holographic Flavored Quark-Gluon Plasmas,''
    [arXiv:1101.3841 [hep-ph]].
  
  
\bibitem{Agranovich}
  V.~M.~Agranovich, Y.~N.~Gartstein,
  ``Spatial dispersion and negative refraction of light,",
  PHYS-USP {\bf 49} (10), 1029-1044 (2006)
\\
  V.~M.~Agranovich, Y.~R.~ Shen, R.~H.~Baughman and A.~A.~ Zakhidov,
  ``Optical bulk and surface waves with negative refraction,"
  Phys.\ Rev.\  B {\bf 69} (2004) 165112
\\
V.~M.~Agranovich and Y~.N.~Gartstein,
``Electrodynamics of metamaterials and the Landau Lifshitz
approach to the magnetic permeability,''
Metamaterials {\bf 3}  19 (2009)

\bibitem{negref}
  V.~G.~Veselago, Sov. Phys. Usp. {\bf 10}, 509 (1968);  \\
http://www.wave-scattering.com/negative.html; \\
D.~R.~Smith, W.~J.~Padilla, J.~Willie, D.~C.~Nemat-Nasser. and S.~Schultz, Phys. Rev. Lett {\bf84}, 4184-4187 (2000);\\
 J.~B.~Pendry, Phys. Rev. Lett. {\bf 85}, 3966 (2000);\\
 R.~A.~Shelby, D.~R.~Smith, S.~Schultz, Science {\bf292}, 77-79 (2001) .
  

\bibitem{Depine}
R.~A.~Depine and A.~Lakhtakia,
 Microwave and Optical Technology Letters, {\bf 41} 315


\bibitem{Landau}
L.~D.~Landau, E.~M.~Lifshitz, "Electrodynamics of continuous media", Oxford, Pergamon
Press, 1984

\bibitem{Dressel}
M.~Dressel and G.~Gruner,
``Electrodynamics of Solids'', Cambridge University Press (2002)


\bibitem{Foster}
  D.~Forster, "Hydrodynamic Fluctuations, Broken Symmetries and Correlation Functions," 
  Benjamin-Cummings, Reading (1975).

\bibitem{KM}
L. P. Kadanoff and P. C. Martin, "Hydrodynamics Equations and Correlation Functions,"
Annals of Physics, 24, 419 (1963).

%\cite{Hartnoll:2007ih}
\bibitem{Hartnoll:2007ih}
  S.~A.~Hartnoll, P.~K.~Kovtun, M.~Muller, S.~Sachdev,
  ``Theory of the Nernst effect near quantum phase transitions in condensed matter, and in dyonic black holes,''
  Phys.\ Rev.\  {\bf B76 } (2007)  144502.
  [arXiv:0706.3215 [cond-mat.str-el]].

%\cite{Herzog:2011ec}
\bibitem{Herzog}
  C.~P.~Herzog, N.~Lisker, P.~Surowka and A.~Yarom,
  ``Transport in holographic superfluids,''
  arXiv:1101.3330 [hep-th].
  %%CITATION = ARXIV:1101.3330;%%

\bibitem{Pekar}
S.~I.~Pekar, Zh. Eksp. Teor. Fiz. 33, 1022 (1957).


\bibitem{ALW}
  A.~Amariti, D.~Forcella, A.~Mariotti,
  ``Additional Light Waves in Hydrodynamics and Holography,''
  [arXiv:1010.1297 [hep-th]].

\bibitem{siani}
  A.~Amariti, D.~Forcella, A.~Mariotti and M.~Siani,
  ``Negative refraction and Superconductivity", appeared today.


\bibitem{Aharony}
O.~Aharony, S.~S.~Gubser, J.~M.~Maldacena, H.~Ooguri, Y.~Oz,
  ``Large N field theories, string theory and gravity,''
  Phys.\ Rept.\  {\bf 323 } (2000)  183-386.
  [hep-th/9905111].



\bibitem{korea}
X.~-H.~Ge, Y.~Matsuo, F.~-W.~Shu, S.~-J.~Sin, T.~Tsukioka,
  ``Density Dependence of Transport Coefficients from Holographic Hydrodynamics,''
  Prog.\ Theor.\ Phys.\  {\bf 120 } (2008)  833-863.
  [arXiv:0806.4460 [hep-th]].
  
  

%\cite{Andronic:2005yp}
%\bibitem{Andronic:2005yp}
 \bibitem{Andronic}
  A.~Andronic, P.~Braun-Munzinger, J.~Stachel,
  ``Hadron production in central nucleus-nucleus collisions at chemical freeze-out,''
  Nucl.\ Phys.\  {\bf A772 } (2006)  167-199.
  [nucl-th/0511071].

\bibitem{Redlich}
%\cite{BraunMunzinger:2003zd}
%\bibitem{BraunMunzinger:2003zd}
  P.~Braun-Munzinger, K.~Redlich, J.~Stachel,
  ``Particle production in heavy ion collisions,''
  In *Hwa, R.C. (ed.) et al.: Quark gluon plasma* 491-599.
  [nucl-th/0304013].


%\cite{CasalderreySolana:2011us}
%\bibitem{CasalderreySolana:2011us}
 \bibitem{Mateos}
  J.~Casalderrey-Solana, H.~Liu, D.~Mateos, K.~Rajagopal, U.~A.~Wiedemann,
  ``Gauge/String Duality, Hot QCD and Heavy Ion Collisions,''
  [arXiv:1101.0618 [hep-th]].
  
  
\bibitem{Stereo}
N.~Liu, H.~Liu, S.~Zhu, H.~Giessen,
``Stereometamaterials",
Nature Photonics 3, 157 - 162 (2009) 

\end{thebibliography}
\end{document}